
\magnification=\magstep1
\rightline{TAUP 2073-93}
\rightline{IASSNS 92/75}
\rightline{July, 1993}
\vskip 3 true cm
\centerline {\bf A Soluble Model for Scattering and Decay}
\centerline {\bf in}
\centerline {\bf Quaternionic Quantum Mechanics I: Decay}
\centerline {L.P. Horwitz\footnote{\S}{Permanent address:
 School of Physics,
Raymond and Beverly Sackler Faculty of Exact Sciences,
 Tel Aviv University,
Ramat Aviv, Israel; also at Department of Physics,
 Bar-Ilan University, Ramat
Gan, Israel.}}
\centerline{School of Natural Sciences}
\centerline{Institute for Advanced Study}
\centerline{Princeton, N.J. 08540}
\vskip 2 true cm
{\it Abstract.\/}  The Lee-Friedrichs model has
 been very useful in the study
of decay-scattering systems in the framework of
 complex quantum mechanics.
Since it is exactly soluble, the analytic
 structure of the amplitudes can be
explicitly studied. It is shown in this paper that a similar model,
which is also exactly soluble, can be
constructed in quaternionic quantum mechanics. The problem of
the decay of an unstable system is treated here.
   The use of the Laplace
transform, involving quaternion-valued
analytic  functions of a variable with values in
a complex subalgebra of the quaternion algebra, makes
the analytic properties of the solution apparent; some
analysis is given of the dominating structure in the analytic
continuation to the
lower half plane.  A study of the
corresponding scattering system will be given in a
succeeding paper.
\vfill
\eject

\noindent {\bf 1. Introduction.}
\smallskip
\par Adler$^{1,2}$ has developed methods
 in time dependent perturbation theory
for the treatment of scattering and decay problems in quaternionic quantum
theory.
  Although the spectrum of the
anti-self-adjoint generator of evolution is quaternion
defined\footnote{\dag}{For an anti-self-adjoint operator $\tilde S$, the
(generalized) eigenvalue equation is
 ${\tilde S} f=fe\lambda, e \in {\rm Im}{\bf H},
e^2 = -1$, where ${\bf H}$ is the
 quaternion algebra spanned over the reals by
$1,i,j,k, ij=k$ and $i^2 = j^2 = k^2 = -1$.  One can choose the quaternion
phase$^1$ of $f$ such that $e=i$ and $\lambda \geq 0$. We call this the
quaternion defined spectrum.} as positive, the resulting formulas for the
scattering matrix display both left
 and right hand cuts.  It has recently been
shown$^3$ that such an anti-self-adjoint
 operator on a quaternionic Hilbert
space has, quite generally, a symmetric
 effective spectrum, and in case the
quaternion defined spectrum is absolutely continuous in $[0,\infty)$, the
effective spectrum is absolutely
 continuous in $(-\infty, \infty)$.  In this
case, a symmetric conjugate operator exists.
  In case the anti-self adjoint
operator is the generator of evolution, the conjugate operator can be
interpreted as a ``time operator''; it was shown$^3$ that there is then no
evident contradiction to the definition of a Lyapunov operator$^4$.
\par  The Lee-Friedrichs model$^5$ has been very fruitful in the study of
decay-scattering systems in the complex Hilbert space$^6$.  The
generalized states (elements of a
 Gel'fand triple, or rigged Hilbert space)
associated with the resonances of the scattering matrix and the reduced
resolvent, for the case in the the unperturbed Hamiltonian is
bounded from below $^7$ as well as for the case for which, as in
a model for the Stark effect, the ``unperturbed'' Hamiltonian
has spectrum $-\infty$ to $\infty$ \  $^8$,
have also been studied
 using this spectral model. For the study
of the
consequences of the symmetrical effective spectrum$^3$ of a quaternionic
quantum theory, as well as for the construction of physical models,
 it would be useful
 to construct an exactly soluble model of
Lee-Friedrichs type in this framework as well.  It is the purpose of this
paper to construct such a model.
\par We first utilize (Section 2) the formulation of the decay
problem suitable for
 application of time-dependent
perturbation theory, following the methods given by
Adler$^{1,2}$ and show that a soluble model can be constructed.
  In Section 3 we
 formulate the problem in terms of the Laplace
transform of the Schr\"odinger evolution.
 The resolvent formalism cannot be
used directly, since there is no natural
 imaginary direction associated with
an anti-self-adjoint operator on a quaternion Hilbert space.   Choosing a
direction arbitrarily, and hence a subalgebra ${\bf C}(1,i)$
of the quaternion algebra ${\bf H}$,
 the Laplace transform (and its inverse) of the
Schr\"odinger equation is well defined.
  In the complex Hilbert space, for a time-independent Hamiltonian, this
procedure is equivalent to the resolvent formalism; for a time dependent
generalization of the Lee-Friedrichs
 model, it has been shown that it provides
a useful basis for a well-controlled perturbation theory$^9$.  For the
quaternionic analog of the Lee-Friedrichs model, we find that the Laplace
transformed amplitude has a structure
 analogous to that of the resolvent form,
but contains a linear combination of a
function analytic in the transform
variable $z\in {\bf C}(1,i)$
 and a function of the negative of its complex
 conjugate $-z^*$, corresponding to the
contributions of the right- and
 left-hand cuts.
The analytic continuation of the amplitude into the lower half plane
is carried out in Section 4, and some results are given for the
structure of the functions there that can dominate the inverse
transform.
In a succeeding paper, some aspects of formal scattering theory
are developed, and
 it is shown that this
model provides a soluble
 scattering problem as well; it therefore describes a
decay-scattering system.
 A summary and conclusions are given in Section 5.
\bigskip
\noindent {\bf 2. Spectral Expansion}
\smallskip
\par In this section, we follow the methods of Adler$^{1,2}$ for the
construction of an effective
 time dependent perturbation theory;
 assuming, as for the Lee-Friedrichs model, that the
potential has no continuum-continuum matrix elements, we obtain a soluble
quaternionic model for a decay-scattering system.
\par In the framework of the theory of Wigner and Weisskopf$^{10}$ , the
component of the wave function evolving under the action of the full
Hamiltonian which remains in the
 initial state corresponds to the probability
amplitude $A(t)$ for the initial state to persist (the survival
amplitude$^{11}$).  Let $\vert \psi_0 \rangle$ be the initial state and
$\tilde H$ be the quaternion
 linear\footnote{\ddag}{There are operators on the
quaternionic Hilbert space with weaker
 linearity properties, such as complex
linear or real linear$^{12}$.  We restrict
 ourselves here to the quaternion
linear case, so that ${\tilde H}(fq) = ({\tilde H}f)
 q$ for all vectors $f$ in
the domain of $\tilde H$.} operator which generates
 the time evolution of the
system.  Then,
$$ A(t) = \langle \psi_0 \vert e^{-{\tilde H}t} \vert \psi_0 \rangle.
\eqno(2.1)$$
\par Let us now suppose that $\tilde H$ has the decomposition
$$ {\tilde H} = {\tilde H}_0 + {\tilde V}, \eqno(2.2)$$
where $\tilde V$ is a small operator which perturbs the eigenstates of
${\tilde H}_0$.  If we consider the eigenstates of ${\tilde H}_0$ as the
quasistable states of the system, $(2.2)$ generates the evolution of an
unstable system.  In particular,
 if $\vert \psi_0 \rangle$ in $(2.1)$ is an
eigenstate of ${\tilde H}_0$, the evolution of $A(t)$ reflects its
instability.
\par Let us assume
 that there is just one eigenstate of ${\tilde
H}_0$, with eigenvalue $E_0$, so that the spectral expansion of $\vert
\psi(t)\rangle = e^{-{\tilde H}t} \vert \psi_0 \rangle $ is
$$ \vert \psi(t)\rangle = \vert \psi_0\rangle e^{-iE_0t} C_0(t) +
\int_0^\infty dE \vert E\rangle e^{-iEt} C_E(t), \eqno(2.3)$$
where
$$ {\tilde H_0} \vert \psi_0 \rangle
 = \vert \psi_0 \rangle iE_0 \eqno(2.4)$$
and
$$ \langle E \vert {\tilde H}_0 \vert f\rangle = iE \langle E \vert f\rangle .
 \eqno(2.5)$$
We take initial values to be
$$ C_0(0) = 1,\qquad  C_E (0) = 0.  \eqno(2.6) $$
\par The transition from the eigenstate
 of ${\tilde H}_0$ to the continuum is
interpreted as the decay of the unstable
 system.  The corresponding amplitude
is
$$ \langle E \vert \psi(t)\rangle = e^{-iEt} C_E(t).
\eqno(2.7)$$
The total transition probability is
$$ \eqalign {\int_0^\infty \vert \langle E \vert \psi(t)\rangle
 \vert ^2 dE &=
\int_0^\infty \vert C_E(t)\vert^2 dE \cr &=1-\vert C_0(t)\vert^2
\cr}\eqno(2.8)$$
\par The Schr\"odinger equation for the evolution,
$$ {\partial \over {\partial t}} \vert \psi(t) \rangle = -{\tilde H} \vert
\psi(t) \rangle, \eqno (2.9)$$
with $(2.3), (2.4)$ and $(2.5)$, becomes
$$\eqalign{\vert \psi_0 \rangle e^{-iE_0t}
 C_0'(t) &+ \int_0^\infty dE\ \vert E
\rangle e^{-iEt} C_E'(t)  \cr
 &= -{\tilde V} \vert \psi_0 \rangle e^{-iE_0t}
C_0(t) - \int_0^\infty dE\ {\tilde V}
 \vert E \rangle e^{-iEt} C_E(t) .\cr}
\eqno (2.10)$$
We take the scalar product in turn
 with $\langle \psi_0 \vert$ and $\langle E
\vert$  to obtain the coupled differential equations$^2$

$$\eqalign{{C_0}'(t)
 &= -e^{iE_0t} \langle \psi_0 \vert {\tilde V}\vert \psi_0
\rangle e^{-iE_0t} C_0 (t) \cr &- \int_0^\infty dE\
 e^{iE_0t} \langle \psi_0 \vert
{\tilde V}\vert E \rangle e^{-iEt} C_E (t), \cr} \eqno(2.11)$$
and

$$\eqalign{ {C_E}'(t)
 &= - e^{iEt} \langle E \vert {\tilde V} \vert \psi_0
\rangle e^{-iE_0t} C_0(t) \cr &- \int_0^\infty dE'\
 e^{iEt} \langle E \vert
{\tilde V} \vert E' \rangle e^{-iE't} C_{E'}(t) . \cr} \eqno(2.12)$$

Integrating these equations over
 a very short interval in $t$, covering $t=0$,
one sees that if we take
 $C_0 (t), C_E (t)$ to be zero for $t \leq 0$, the
conditions $(2.6)$ can be realized
 in the form $C_0(0_+) = 1, C_E(0_+)= 0,$ by
adding a term $\delta(t)$ to the right hand side$^2$
 of $(2.11)$.

\par We now assume a condition on
 $\tilde V$ parallel to that of the Lee-Friedrichs
 model in the complex quantum theory, i.e., that
$$ \langle E \vert {\tilde V}\vert E' \rangle = 0. \eqno(2.13)$$
The imaginary complex part of $\langle \psi_0 \vert {\tilde V}\vert \psi_0
\rangle$,  as for the Lee-Friedrichs model of
 the complex quantum theory, contributes a level shift (i.e., it
appears only in the combination $E_0 + v$, where $iv$ is the complex
part of ${\tilde V}$[{\it cf.}$(2.16)$]), but
the imaginary quaternionic part contributes to the
 quaternionic phase of the (sub-asymptotic)
decay amplitude\footnote{*}{ Since
we have already chosen the quaternionic phase of $\vert \psi_0 \rangle$
so that the eigenvalue of ${\tilde H}_0$ is $iE$, with $E>0$, there is
no further freedom to reray $\langle \psi_0 \vert {\tilde V}\vert
\psi_0 \rangle $ so that it lies in ${\bf C}(1,i)$.}.
  We shall retain the term $\langle \psi_0 \vert {\tilde V}\vert \psi_0
\rangle$
here to display its effect.  The condition $(2.13)$ is essential to the
solubility of the model; it corresponds to neglecting final state
interactions, often (as in radiative processes)
 a good approximation to the
physical problem in the complex quantum theory.
  Eqs. $(2.11), (2.12)$ then
become
$$ \eqalign{{C_0}'(t) &= -e^{iE_0t} \langle \psi_0
 \vert {\tilde V} \vert \psi_0
\rangle e^{-iE_0t} C_0(t) \cr &- \int_0^\infty dE\
 e^{iE_0t} \langle \psi_0 \vert
{\tilde V} \vert E \rangle e^{-iEt} C_E (t)
 + \delta (t) \cr} \eqno(2.14) $$
and
$${C_E}'(t) = - e^{iEt} \langle E \vert {\tilde V} \vert \psi_0 \rangle
e^{-iE_0 t}C_0 (t).  \eqno(2.15)$$
To make use of the methods of complex analysis, we decompose all of the
quaternion-valued functions in $(2.14), (2.15)$ into their symplectic
components$^{2,12,13}$, i.e.,
$$ \eqalign{C_0 (t) &= C_{0\alpha} (t) + j C_{0\beta}(t) \cr
        C_E(t) &= C_{E\alpha}(t) + j C_{E\beta}(t) \cr
  \langle E \vert {\tilde V}
   \vert \psi_0 \rangle &= - \langle \psi_0 \vert
{\tilde V} \vert E \rangle^* \equiv  V_{E\alpha} + j V_{E\beta}, \cr
\langle \psi_0 \vert {\tilde V} \vert \psi_0 \rangle &= iv +ju, \cr}
\eqno(2.16)$$
where, since $\langle \psi_0 \vert {\tilde V} \vert \psi_0 \rangle ^* = -
\langle \psi_0 \vert {\tilde V}
 \vert \psi_0 \rangle $, $v$ is real, and  $u \in {\bf C}(1,i)$;
the asterisk is used here to denote the quaternion conjugation (it then
corresponds to complex conjugation on the complex subalgebra ${\bf
C}(1,i)$).
  Then, Eqs. $(2.14)$ and $(2.15)$ become

$$ \eqalign{ C_{0\alpha}'(t) &= -iv C_{0\alpha}(t) + u^* e^{2iE_0t}
C_{0\beta}(t) \cr
&+\int_0^\infty dE\ e^{i(E_0-E)t} V_{E\alpha}^* C_{E\alpha}(t)  \cr
&+\int_0^\infty dE\ e^{i(E_0 + E)t} V_{E\beta}^* C_{E\beta}(t) +\delta(t), \cr
C_{0\beta}'(t) &= -e^{-2iE_0t} u C_{0\alpha}(t) + iv C_{0\beta}(t) \cr
&-\int_0^\infty dE\ e^{-i(E_0+E)t} V_{E\beta} C_{E\alpha}(t)  \cr
&+ \int_0^\infty dE\ e^{-i(E_0 - E)t} V_{E\alpha} C_{E\beta}(t)  \cr}
\eqno(2.17)$$
and

$$\eqalign{ C_{E\alpha}'(t)
 &= -e^{i(E-E_0)t} V_{E\alpha} C_{0\alpha} (t) +
e^{i(E+E_0)t} V_{E\beta}^*C_{0\beta}(t) \cr
C_{E\beta}'(t) &= -e^{-i(E+E_0)t}
 V_{E\beta}C_{0 \alpha}(t) - e^{-i(E-E_0)t}
V_{E\alpha}^* C_{0\beta(t)}. \cr}\eqno(2.18)$$
We now define the Fourier transforms $^2$
$$\eqalign{ C_{0\alpha} (t) &= - {1 \over 2\pi i}\int_{-\infty}^\infty
d\omega e^{i(E_0-\omega)t}
 C_{0\alpha}(\omega) \cr C_{0\beta}(t) &= {1 \over
2\pi i} \int_{-\infty}^\infty
 d\omega e^{-i(E_0+\omega)t}C_{0\beta}(\omega)
\cr C_{E\alpha}(t)
 &= -{1\over 2\pi i}\int_{-\infty}^\infty d\omega e^{i(E-
\omega)t} C_{E\alpha} (\omega) \cr C_{E\beta} (t) &= {1 \over 2\pi i}
\int_{-\infty}^\infty d\omega e^{-i(E+\omega)t} C_{E\beta}(\omega) \cr
i\delta(t) &= -{1 \over 2\pi i} \int_{-\infty}^\infty d\omega
e^{i(E_0-\omega)t}  \cr} \eqno(2.19) $$

Note that it follows from the conditions
$C_0(t), C_E(t) = 0$ for $t \leq 0$, that the complex-valued functions
$C_{0\alpha}(\omega), C_{0\beta}(\omega), C_{E\alpha}(\omega),
C_{E\beta}(\omega)$ are analytic in the upper half complex $\omega$-plane
(upper half plane analyticity
 is equivalent to the condition that $C_0(t), C_E(t) $
 vanish for $t\leq 0$).

\par With the definitions $(2.19)$, Eqs.
$(2.17)$ and $(2.18)$ become
$$ \eqalign{-i &= -i(\omega - E_0 -v)C_{0\alpha}(\omega) + u^* C_{0\beta}
(\omega) \cr &+ \int_0^\infty dE
 \bigl({-V^*_{E\alpha} C_{E\alpha}(\omega) +
V_{E\beta}^*C_{E\beta}(\omega)\bigr)}\cr 0&= i(\omega + E_0 +v)
C_{0\beta}(\omega) + u C_{0\alpha}(\omega) \cr
 &+ \int_0^\infty dE \bigl({
V_{E\beta} C_{E\alpha}(\omega)
 + V_{E\alpha} C_{E\beta}(\omega) \bigr)}, \cr}
\eqno(2.20) $$
and

$$ \eqalign{ C_{E\alpha} (\omega) &= {i \over {E-\omega}} \bigl(
V_{E\alpha}C_{0\alpha} (\omega) +
 V_{E\beta}^* C_{0\beta} (\omega) \bigr) \cr
C_{E\beta}(\omega) &= {i \over {E+\omega}} \bigl( V_{E\beta}
C_{0\alpha} (\omega) - V_{E\alpha}^* C_{0\beta} (\omega) \bigr), \cr}
\eqno(2.21)$$
where the division is well-defined for $\omega$ in the upper half plane.
Substituting $(2.21)$ into $(2.20)$, one obtains the closed form

$$ \eqalign{ h(\omega) C_{0\alpha}
 (\omega) + g(-\omega^*)^* C_{0\beta}(\omega) &=1\cr
g(\omega) C_{0\alpha}(\omega)
 - h(-\omega^*)^* C_{0\beta} (\omega) &=0, \cr}
\eqno(2.22)$$
where

 $$\eqalign{ h(\omega) &= \omega -E_0 -v + \int_0^\infty dE \biggl( {\vert
V_{E\alpha} \vert^2 \over {E-\omega}} -{\vert V_{E\beta}\vert^2 \over
{E+\omega}} \biggr), \cr
g(\omega) &= -iu + \int_0^\infty dE\
V_{E\alpha}V_{E\beta} \biggl( {1 \over {E-\omega}} + {1 \over {E+\omega}}
\biggr). \cr} \eqno(2.23)$$
Note that $h(\omega)$ and $g(\omega)$ are analytic in the upper half
plane; it therefore follows that $h(-\omega^*)^*$
 and $g(-\omega^*)^*$ are also analytic
  functions of $\omega$ in the
upper half plane.
\par  The Eqs. $(2.22)$ provide a closed
 solution for the coefficients
$C_{0\alpha}(\omega), C_{0\beta}(\omega)$ in the domain of analyticity Im
$\omega >0$. The time dependent
 coefficients $C_{0\alpha}(t), C_{0\beta}(t)$
can then be obtained by inverting
 the Fourier transform, integrating on a line
infinitesimally above the real axis in the $\omega$ plane (implicit in the
relations $(2.19)$).
We wish to emphasize that unlike the Lee-Friedrichs
model with semi-bounded unperturbed
 Hamiltonian, the cut is not just a half
line. In accordance
 with what we have found for the effective
spectrum of an anti-self-adjoint
 operator in ref. 3, these results reflect the
presence of both a left and right hand cut$^{14}$. The integrals can
 be estimated by lowering the contour into
the lower half plane (the second Riemann
 sheet)$^8$, where the analytic continuation of the first
sheet functions are well-defined, to obtain the contribution of leading
singularities.  We study this analytic continuation in Section 4.
\bigskip
\noindent {\bf 3. Laplace Transforms}
\smallskip
\par
In this section we study the
 Laplace transform of the amplitude $A(t)$
defined in $(2.1)$. As we have pointed out above, the direct
application of resolvent methods is complicated by the quaternion
structure of the Hilbert space.  However, the Laplace transform
nevertheless brings the equations to a form which is similar
to that of the resolvent method in the complex Hilbert
space.

 \par
The amplitude $A(t)$ satisfies the equation

  $$ \eqalign{- {\partial \over \partial t} A(t) &= \langle \psi_0
  \vert ({\tilde H} + {\tilde V})\vert \psi(t)\rangle \cr
  &= iE_0 A(t) + \langle \psi_0 \vert {\tilde V} \vert \psi(t)
  \rangle ,\cr} \eqno(3.1) $$
where we have used $(2.4)$.
  \par

Inserting the completeness relation
$$ 1 = \vert \psi_0 \rangle \langle \psi_0 \vert + \int_0^\infty dE\
\vert E \rangle \langle E \vert $$
between ${\tilde V}$ and $\vert \psi(t) \rangle$, this is
$$\eqalign{-{\partial \over \partial t} A(t) &= iE_0 A(t)
+ \int_0^\infty dE \langle \psi_0 \vert {\tilde V} \vert
E \rangle A_E(t), \cr &+ \langle \psi_0 \vert {\tilde V}
\vert \psi_0 \rangle A(t), \cr} \eqno(3.2)$$
where

$$ A_E(t) = \langle E \vert \psi(t) \rangle .\eqno(3.3)$$
The amplitude $A_E(t)$ satisfies

$$\eqalign{ -{\partial \over \partial t} A_E(t) &= \langle E
\vert {\tilde H}_0 + {\tilde V} \vert \psi(t) \rangle \cr
&= iE A_E(t) + \langle E \vert {\tilde V} \vert \psi_0
\rangle A(t) , \cr} \eqno(3.4)$$
where we have used the completeness relation again, taking
into account the restriction $(2.13)$
on the matrix elements that constitute the soluble model, i.e.,
that ${\tilde V}$ connects the continuum only to the discrete
eigenstate.  We
now define the Laplace transform:

$$ {\hat A}(z)= \int_0^\infty dt\ e^{izt} A(t) , \eqno(3.5) $$
and a similar expression for ${\hat A}_E(z)$.  Clearly,
${\hat A}(z)$ exists and is analytic in the upper half
$z$-plane, since by the Schwarz inequality$^2$ $\vert
A(t) \vert \leq \Vert \psi_0 \Vert \Vert e^{-{\tilde H}t}\psi_0
\Vert = \Vert \psi_0 \Vert^2$, i.e., $A(t)$ is bounded.
The upper half plane analyticity of ${\hat A}_E(z)$ follows
from that of ${\hat A}(z)$, as we shall see below.
\par  The inverse Laplace transform follows as for the usual
complex function space, since the quaternion structure
of $A(t)$ does not affect the left transform of
${\hat A}(z)$.  For $C$ a line from $ -\infty +i\epsilon$
to $+\infty +i\epsilon$,
above the real axis in the $z$-plane, we assert that
$$ A(t) = {1 \over 2\pi} \int_C dz\  e^{-izt} {\hat A}(z) .
\eqno(3.6)$$
For $t < 0\  ( {\hat A}(z) \rightarrow 0\ {\rm for\ Im}z \rightarrow
i\infty)$, the contour can be closed in the upper half-
plane, so that, by upper half-plane analyticity, $A(t) =0$.
For $t \geq 0$,

$$ \eqalign{A(t) &={1 \over 2\pi}
 \int_C dz e^{-izt} \int_0^\infty dt'
e^{izt'}A(t') \cr &= {1 \over 2\pi} \int_0^\infty dt' \
e^{\epsilon (t-t')} \int_{-\infty}^\infty dx\ e^{-ix(t-t')} A(t') ,\cr}
\eqno(3.7)$$
\noindent and, since $\int_{-\infty}^\infty dx\ e^{-ix(t-t')}
= 2\pi \delta(t-t'),$ the assertion is verified.

\par
The Laplace transform of Eq. $(3.2)$ is\footnote{$\natural$}{Note that
the first term on the left hand side is due to integration by parts,
i.e.,
$$\int_0^\infty dt\ e^{izt}\bigl(-{\partial \over \partial t} A(t)
\bigr) = -e^{izt}A(t)\vert_0^\infty + iz \int_0^\infty dt\ e^{izt}A(t) =
1 + iz {\hat A}(z)$$.}
$$\eqalign{1 + iz {\hat A}(z) &=  iE_0{\hat A}(z)
- \int_0^\infty dE\ V_{E\alpha}^* {\hat A}_E(z) \cr
&+ j\int_0^\infty dE\ V_{E\beta} {\hat A}_E(-z^*)
+ iv{\hat A}(z) + ju{\hat A}(-z^*),\cr} \eqno(3.8)$$
and of Eq. $(3.4)$,
$$ iz {\hat A}_E(z)= iE {\hat A}_E(z) + V_{E\alpha} {\hat A}(z)
+ j V_{E\beta}{\hat A}(-z^*).\eqno(3.9)$$
These formulas reduce to the results for the complex case if
$V_{E\beta}$ and $u$ are zero.  The contribution of $\langle \psi_0
\vert {\tilde V}\vert \psi_0 \rangle$ is then,
as is clear from $(3.8)$, just an energy shift.
\par In terms of the components ${\hat A}_\alpha (z)$,
 ${\hat A}_\beta (z)$, ${\hat A}_{E\alpha}(z)$ and ${\hat
A}_{E\alpha}(z)$,
where (we shall see below that the $\alpha$ components are analytic
functions of $z$, and the $\beta$ components,
 in spite of the notation,
are analytic functions of $z^*$)
$$\eqalign{{\hat A}(z)&= {\hat A}_\alpha (z) + j {\hat A}_\beta (z),\cr
{\hat A}_E(z)&= {\hat A}_{E\alpha} (z) + j {\hat A}_{E\beta} (z), \cr}
\eqno(3.10)$$
\noindent one obtains
$$\eqalign{ 1 + iz{\hat A}_\alpha(z) &= iE_0 {\hat A}_\alpha (z)
-\int_0^\infty dE\ V_{E\alpha}^* {\hat A}_{E\alpha}(z) \cr
&- \int_0^\infty dE\ V_{E\beta}^* {\hat A}_{E\beta}(-z^*) +iv{\hat
A}_\alpha(z) -u^* {\hat A}_\beta(-z^*), \cr
-iz^* {\hat A}_\beta(z) &= -iE_0 {\hat A}_\beta (z) -
\int_0^\infty dE\ V_{E\alpha} {\hat A}_{E\beta}(z) \cr
&+ \int_0^\infty dE\ V_{E\beta} {\hat A}_{E\alpha} (-z^*) -iv{\hat
A}_\beta(z) + u {\hat A}_\alpha (-z^*) ,\cr} \eqno(3.11)$$
and
$$ \eqalign{ iz{\hat A}_{E\alpha}(z) &= iE {\hat A}_{E\alpha}(z)
+ V_{E\alpha} {\hat A}_\alpha (z) - V_{E\beta}^* {\hat A}_\beta
(-z^*) \cr -iz^* {\hat A}_{E\beta}(z) &= -iE {\hat A}_{E\beta}(z)
+ V_{E\alpha}^* {\hat A}_\beta(z) + V_{E\beta} {\hat A}_\alpha
(-z^*) \cr} \eqno(3.12)$$
\par We now study the relation between the amplitudes ${\hat A}(z),
\ {\hat A}_E(z)\  {\rm and}\  C_0(z),\  C_E(z)$,
 and show that the equations
$(3.8)$ and  $(3.9)$ are equivalent to $(2.20)$ and $(2.21)$.
\par The transforms defined in $(2.19)$ have the property
$$C_{E\alpha}(t) + j C_{E\beta}(t) = -{1 \over 2\pi i}
e^{iEt} \int_{-\infty}^\infty d\omega \{e^{-i\omega t}
C_{E\alpha}(\omega) + e^{i\omega t} j C_{E\beta}(\omega) \},
\eqno(3.13)$$
a relation valid for the $C_{E_0}(t)$ coefficients as
well.  The transforms act with opposite sign of the frequency
on the $\alpha$ and $\beta$ parts of the $C$-amplitudes;
they are, however, analytic in the upper half $\omega$-plane by
construction. On the other hand, the transform $(3.5)$,
 also analytic in the upper
half $z$-plane, for
$$A(t) = A_\alpha(t) + j A_\beta(t),  \eqno(3.14) $$
has the form
$${\hat A}(z) = \int_0^\infty dt\ e^{izt}A_\alpha (t)
+j \int_0^\infty dt\ e^{-iz^*t} A_\beta (t), \eqno(3.15)$$
and hence, with $(3.10)$, we see that
$$\eqalign{ {\hat A}_\alpha (z) &=
 \int_0^\infty dt\ e^{izt}A_\alpha (t)  \cr {\hat A}_\beta (z)
&= \int_0^\infty dt\ e^{-iz^*t}A_\beta (t)  . \cr} \eqno(3.16)$$

The second of $(3.16)$ is an analytic function
of $z^*$ in the upper half $z$-plane (for which $-z^*$ is
also in the upper half plane).
\par This form is, in fact a general one$^{15}$ for quaternion-valued
left analytic functions of a variable in the
 complex subalgebra ${\bf C}(1,i)$.  To see this,
consider the Cauchy-Riemann relations for the existence of a left-acting
derivative with respect to a variable in the complex subalgebra for

$$ \eqalign{f(z) &= f_\alpha(z) + j f_\beta(z)\cr
&= u_\alpha + iv_\alpha + j ( u_\beta + i v_\beta).\cr}$$

Then,

$$ \eqalign{{\partial f \over \partial x } &=
{\partial u_\alpha
\over \partial x} + i {\partial v_\alpha \over \partial x}
+ j \biggl( {\partial u_\beta \over \partial x} + i {\partial
v_\beta \over \partial x} \biggr) \cr
{1 \over i} {\partial f \over \partial y} &=
{1 \over i } {\partial u_\alpha \over \partial y} +
{\partial v_\alpha \over \partial y} - j \biggl(
{1 \over i}{\partial u_\beta \over \partial y} +
{\partial v_\beta \over \partial y} \biggr) \cr}$$
so that if the left derivative of $f(z)$ exists,

$$ {\partial u_\alpha \over \partial x} =
{\partial v_\alpha \over \partial y} \qquad {\partial v_\alpha
\over \partial x} = -{\partial u_\alpha \over \partial y}
\eqno(3.17)$$
and

$$ {\partial u_\beta \over \partial x} = -{\partial v_\beta
\over \partial y} \qquad {\partial v_\beta \over \partial
x} = {\partial u_\beta \over \partial y}. \eqno(3.18)$$

The relations $(3.18)$ correspond to anti-Cauchy-Riemann
conditions, implying that $f_\beta$ is an analytic function
of $z^*$ [note that with the convention we have used, i.e, of writing
$j$ to the left of the $\beta$ part of the function, {\it right}
analyticity results in normal Cauchy-Riemann conditions for both parts;
later we shall study a function with $j$ appearing to the right of
the $\beta$-part, for which the normal Cauchy-Riemann conditions
for left analyticity apply to both parts].
\par From $(2.3)$, we have the relations

$$ \eqalign {A(t) &= e^{-iE_0 t} C_0(t),\cr
A_E(t) &= e^{-iEt} C_E(t).\cr} \eqno(3.19)$$
It then follows from $(2.19)$ that the Laplace transform
 of $A(t)$ is

$$ {\hat A}(z) = {1 \over 2 \pi} \int_{-\infty}^\infty d\omega
\biggl( {C_{0\alpha} (\omega) \over {\omega - z}}-
j {C_{0\beta} (\omega) \over {\omega - (-z^*)}} \biggr).\eqno(3.20)$$

The functions $C_{0\alpha} (\omega), C_{0\beta}(\omega)$
are analytic in the upper half $\omega$-plane, and for
smooth functions $C_{0\alpha}(t), C_{0\beta}(t)$, vanish
for ${\rm Im}\ \omega \rightarrow \infty$.  Hence, one can
complete the contour in $(3.20)$ to obtain

$$\eqalign{ {\hat A}_\alpha(z) &= i C_{0\alpha}(z)\cr
{\hat A}_\beta(z) &= -i C_{0\beta}(-z^*), \cr} \eqno(3.21)$$
consistent with the upper half plane left analyticity of ${\hat A}(z)$.
The functions ${\hat A}_{E\alpha}(z),
{\hat A}_{E\beta}(z)$ are related in the same way to $C_{E\alpha}(z),
C_{E\beta}(-z^*)$.
\par Substituting the relations $(3.21)$ into $(3.11)$ and $(3.12)$,
one obtains
$$ \eqalign{-i &= -i(z - E_0 -v)C_{0\alpha}(z) + u^* C_{0\beta}
(z) \cr &+ \int_0^\infty dE
 \bigl(-V^*_{E\alpha} C_{E\alpha}(z) +
V_{E\beta}^*C_{E\beta}(z)\bigr)\cr 0&= i(E_0 +v -z^*)
C_{0\beta}(-z^*) + u C_{0\alpha}(-z^*) \cr
 &+ \int_0^\infty dE \bigl(
V_{E\beta} C_{E\alpha}(-z^*)
 + V_{E\alpha} C_{E\beta}(-z^*) \bigr), \cr}
\eqno(3.22) $$
and

$$ \eqalign{ C_{E\alpha} (z) &= {i \over {E-z}} \bigl(
V_{E\alpha}C_{0\alpha} (z) +
 V_{E\beta}^* C_{0\beta} (z) \bigr) \cr
C_{E\beta}(-z^*) &= {i \over {E-z^*}} \bigl(  V_{E\beta} C_{0\alpha}
 (-z^*)- V_{E\alpha}^*
C_{0\beta} (-z^*)  \bigr). \cr}
\eqno(3.23)$$
The replacement of $-z^*$ by $z$ in the second of each of
 $(3.22)$ and $(3.23)$ corresponds to an analytic continuation
of these equations inside the upper half plane, where the functions
are analytic.  This replacement brings these expressions into
correspondence with $(2.20)$ and $(2.21)$.
\par Let us now return to the quaternionic equations $(3.8)$ and
$(3.9)$. Substituting the second into the first to eliminate the
amplitude ${\hat A}_E(z)$, we obtain
$$ 1 + ih(z){\hat A}(z) = -ijg(-z^*){\hat A}(-z^*), \eqno(3.24)$$
where $h$ and $g$ are the functions defined in $(2.23)$. Note that
although it appears that $g(z)$ , as defined by $(2.23)$,
is an even function of its
argument, it is , in fact, defined in the upper half plane, and has a
cut along the real axis, so that the formal reflection of $z$ takes
the function out of its original domain of definition.
The domain of definition of the function can be extended
to the lower half plane by analytic
continuaton, which we shall discuss below.
  The same remark is
applicable to the function $h(z)$, which appears,
in $(2.23)$,
 to have a trivial
reality property, i.e., that $h(z)^* = h(z^*)$; this operation,
however, also shifts the variable
from the upper to the lower half plane.  The value of the
extension of this function to the lower half plane also involves
analytic continuation, which we shall discuss below.
\par  If $V_{E\beta}$ vanishes, then $g(z)$ vanishes, and
the result $(3.24)$ coincides precisely with the
well-known result for complex quantum theory, usually obtained by
means of the method of resolvents$^{6,7}$.  The quaternionic part
of the transition matrix element couples the amplitude to its reflection
in the imaginary axis of the complex energy plane, and corresponds to
the effect of the negative
 energy part of the effective spectrum of the
anti-self-adjoint Hamiltonian$^5$.
\par Since $h(z)$, $g(z)$ and ${\hat A}(z)$ are analytic in the upper
half plane, Eq.$(3.24)$ can be shifted to the point $-z^*$, providing
the second coupled equation
$$ 1 + ih(-z^*) {\hat A}(-z^*)= -ijg(z) {\hat A}(z). \eqno(3.25)$$
Solving for ${\hat A}(-z^*)$ from the $(3.25)$, and substituting into
$(3.24)$, one obtains the relation
$$ H(z) {\hat A}(z) = iG(z),\eqno(3.26)$$
where
$$H(z) = \overline{h}(z)h(z) + \overline{g}(z)g(z) \eqno(3.27)$$
and
$$G(z) = \overline{h}(z) + \overline{g}(z)j, \eqno(3.28)$$
and we have used the notation
$$ \overline{h}(z) \equiv h(-z^*)^*, \qquad \overline{g}(z)
\equiv g(-z^*)^*  .     \eqno(3.29)$$
The functions $\overline{h}(z)$ and $\overline{g}(z)$
are analytic functions of $z$ if $h(z)$ and $g(z)$ are;
hence $H(z)$ and
$G(z)$ are analytic functions of $z$ as well.
The function $g(z)$ contains the complex quantity $u$, the quaternionic
part of  $\langle \psi_0 \vert {\tilde V} \vert \psi_0 \rangle$, which
therefore
contributes to the  quaternionic part of the decay amplitude $A(t)$.
\bigskip
\noindent
{\bf 4. Analytic Continuation}
\smallskip
\par The inverse of the Laplace transform $(3.6)$ can be
 well-approximated when there are
  singularities in the analytic continuation
of the amplitude ${\hat A}(z)$ into the lower half plane, close to the
real axis, by deforming the countour of the integration $(3.6)$ below the
real axis.  Since, in Eq. $(3.26)$, $G(z)$ depends on functions which
can be taken to have regular analytic continuation to the lower half
plane, such singularities arise, in general, from zeros of $H(z)$.
 \par We shall study in this section the analytic properties of $h(z)$ and
$g(z)$ to see how zeros of $H(z)$ could arise. To do this, it is
convenient to put these functions into a somewhat simpler form by
defining numerator functions in the integrals that can be continued
from the interval $[0,\infty)$ to $(-\infty, \infty)$. As pointed out
in ref. 5, the effective negative spectrum of ${\tilde H}_0$ can be
represented by the identification
(we use a double bar on the Dirac kets to indicate the spectral
representation for which the argument may be positive or negative), for
$E>0$,
$$ \langle -E \| = -j\langle E \vert . \eqno(4.1)$$
The matrix element
$$ V_E = \langle E \vert {\tilde V} \vert \psi_0 \rangle \eqno(4.2)$$
then has the extension
$$ V_{-E} = \langle -E \| {\tilde V} \vert \psi_0 \rangle =
-j \langle E \vert {\tilde V} \vert \psi_0 \rangle,
\eqno(4.3)$$
so that, for $E>0$,
$$            V_{-E\alpha} = V_{E\beta} $$
and
$$            V_{-E\beta} = -V_{E\alpha}.  \eqno(4.4)$$
Since the limit of  $V_{-E\alpha}$ as $E \rightarrow 0_+$ is not
necessarily equal to the limit of
 $V_{E\alpha}$ as $E \rightarrow 0_+$
(we do not require continuity across zero), $(4.4)$ does not
place any restriction on the relation between $V_{E\alpha}$ and
$V_{E\beta}$.
\par With these identifications, it follows that
$$ \eqalign {\int_{-\infty}^0 dE\ {\vert \langle E\| {\tilde V}
\vert \psi_0 \rangle_\alpha \vert^2 \over {E-z}} &=
\int_\infty^0 dE\ {\vert \langle -E \| {\tilde V}\vert
\psi_0 \rangle_\alpha \vert^2 \over {E+z}} \cr &=
- \int_0^\infty dE\ {\vert V_{E\beta}\vert^2 \over {E+z}},\cr}
\eqno(4.5)$$
so that we can write
$$h(z) = z-E_0-v + \int_{-\infty}^\infty dE\ {X(E) \over {E-z} },
\eqno(4.6)$$
where
$$ X(E) = \vert V_{E\alpha}\vert^2, \eqno(4.7)$$
for $E \in (-\infty,\infty)$.
\par In a similar way, since (by $(4.4)$)
$$ V_{-E\alpha}V_{-E\beta} = - V_{E\beta} V_{E\alpha},$$
we may write
$$ g(z) = \int_{-\infty}^\infty dE\ {Y(E)\over {E-z}} -iu ,\eqno(4.8)$$
where
$$ Y(E) = -Y(-E) = V_{E\alpha}V_{E\beta}. \eqno(4.9)$$
\par Taking the imaginary part of $(4.6)$, one obtains
$$ {\rm Im}h(z) = {\rm Im}z \biggl( 1 + \int_{-\infty}^\infty dE\
{X(E) \over \vert E-z\vert^2} \biggr), \eqno(4.10)$$
so that there are no zeros of $h(z)$ for $z$
 in the upper half plane
 (this is therefore true of $\overline{h}(z)$
as well).
\par  Taking $z$ onto the real axis from above, one finds,
from $(2.23)$ with $z$ in place of $\omega$,
$$ h(\lambda) = \cases{ \lambda -E_0 -v +
\pi i  \vert V_{\lambda\alpha}\vert^2 +
P\int_0^\infty dE\
\bigg( {|V_{E\alpha}|^2 \over {E-\lambda}} - {|V_{E\beta}|^2 \over {E+\lambda}}
\biggr), &for $z \rightarrow \lambda >0$;\cr
-\lambda -E_0 - v +
\pi i  | V_{\lambda\beta}|^2
 +P\int_0^\infty dE\
\bigg( {|V_{E\alpha}|^2 \over {E+\lambda}}
 - {|V_{E\beta}|^2 \over {E-\lambda}}
\biggr), &for $z\rightarrow -\lambda
<0$,\cr} \eqno(4.11)$$
where the principal part restriction is not necessary in one term of
each of the integrals.
Since $\vert V_{-\lambda\alpha}\vert^2 = \vert V_{\lambda\beta}\vert^2$,
this result follows, equivalently,  from $(4.6)$ as well.
 If we now assume that we have chosen a model (specified by $v$, $u$ and
the functions $V_{E\alpha}$ and $V_{E\beta}$) for which
$X(E)$ is the boundary value of a function $X(z)$
analytic in the lower half
plane in a large enough domain
 below the real axis to cover the analytic structure
 we must study, $h(z)$ has the
analytic continuation in the lower half plane
$$ h^{II}(z) = z - E_0 - v + \int_{-\infty}^\infty dE\ {X(E) \over
{E-z}} + 2\pi i X(z). \eqno(4.12)$$
We see that taking the limit of $h^{II}(z)$ onto the real axis from
below, one obtains exactly $(4.11)$; a term $-\pi i
|V_{\lambda\alpha}|^2$ for $ z \rightarrow \lambda >0$, or $-\pi i
|V_{\lambda\beta}|^2$ for $z \rightarrow -\lambda <0$, emerges from
the integral, cancelling half of the additional term $2 \pi i X(z)$.
Hence $h(z)$ and $h^{II}(z)$ are
 functions analytic, respectively, in some domain in
the upper half plane and in some domain of the lower half plane, with
a common segment of the real axis on which they approach the same
values; a simple application of the Cauchy theorem demonstrates
that they form one analytic function, and hence are unique analytic
continuations of each other.
It then follows that
$$ {\rm Im} h^{II}(z) = 2\pi {\rm Re}X(z) +
 {\rm Im}z \biggl( 1 + \int_{-\infty}^\infty
dE\ {X(E) \over \vert E-z\vert^2} \biggr), \eqno(4.13)$$
which, if ${\rm Re}X(z) >0$, to be expected if $z$ is close
to the real axis, can vanish for ${\rm Im}z$ negative.
We shall assume for the remainder of the analysis, for simplicity, that
$X(E) \ll E_0$ for every $E$, so that the assumption
that the root of $h^{II}(z)$ has small imaginary part is justified.
\par In a similar way, one finds the analytic continuation of
$g(z)$ to be
$$ g^{II}(z) = \int_{-\infty}^\infty dE\ {Y(E) \over {E-z}}
+ 2\pi i Y(z) -iu, \eqno(4.14)$$
where we have assumed $Y(E)$ to be the boundary value of an
analytic function $Y(z)$, analytic in the same domain as $X(z)$.
There are no model independent statements we can make about $g^{II}(z)$
since $Y(E)$ has no definite sign or reality properties; we shall
therefore further
assume, for the purposes of our present discussion, that
$V_{E\beta} \ll V_{E\alpha}$, and that $u$ is also small, so that
the principal contribution to the contour integral $(3.6)$ comes from
values of $z$ close to roots of
 the analytic continuation of
 the first term, $\overline{h}(z)h(z)$, in
the expression $(3.27)$ for $H(z)$.  With this assumption,
only a small adjustment in the value $z$ of a root of this term
is required to cancel the contribution
of the second term as well.  We therefore study the zeros of
 $\overline{h}(z)h(z)$.
\par It follows from $(4.12)$ that the root $\zeta$ must satisfy
$$\eqalign{ {\rm Re} h^{II}(\zeta) = 0 &=
{\rm Re}\zeta \biggl( 1 -\int_{-\infty}^\infty
dE\ {X(E) \over |E-\zeta|^2} \biggr)\cr
 &- E_0 -v + \int_{-\infty}^\infty
dE\ E {X(E) \over |E-\zeta|^2} -2 \pi
 {\rm Im} X(\zeta),\cr} \eqno(4.15)$$
so that a zero of the real part can only occur at
$$ \omega_0={\rm Re} \zeta \cong E_0 + v. \eqno(4.16)$$
If $|v| < E_0$, it
 follows from $(4.12)$ and $(4.13)$ (and the
assumption that $X(E)$ is small) that a zero can only form
 in $h^{II}(z)$ below the positive real axis.  The
vanishing of the imaginary part then requires, according to $(4.13)$,
that
$$ {\rm Im} \zeta \cong - \pi {\rm Re}X(\zeta) \cong - \pi |V_{\omega_0
 \alpha}|^2,\eqno(4.17)$$
where we understand that $\omega_0 \cong E_0 + v $
 is positive, and we have approximated
  $X(\zeta)$ by its value on the real axis.
This result is obtained from $(4.13)$ by noting that the integral, for
small values of ${\rm Im}z$, contains an approximation to a
representation of the $\delta$-function, i.e.(for $\epsilon >0$),
$$ \lim_{\epsilon \rightarrow 0}{ \epsilon \over {x^2 + \epsilon^2}}
 = \pi \delta(x).$$
Hence,
$$ {|{\rm Im}z| \over \vert E-z\vert^2}
 \cong  \pi \delta (E-{\rm Re}z) ;$$
substituting this back into $(4.13)$,
 with ${\rm Im}h^{II}(z) = 0,$ one
obtains $(4.17)$.  Using the same approximation, this result may be
obtained directly from $(4.12)$; the integral, for small negative ${\rm
Im}z$ contributes an imaginary  term
 which is approximately $-\pi i X(\omega_0)$.
\par We now show that there is a {\it second} leading contribution
to the inverse transform $(3.6)$ when the contour is deformed
into the lower half plane.  The function
$$ \overline{h}(z) \equiv h(-z^*)^* =
-z -E_0 -v + \int_{-\infty}^\infty dE \ {X(E) \over {E+z}}
\eqno(4.18)$$
is analytic in the upper half plane and (following the argument
leading to $(4.10)$ for $h(z)$) has no zeros for ${\rm Im}z \not= 0$;
its analytic continuation to the lower half $z$-plane, obtained
following the procedure resulting in $(4.12)$, is
$$ \overline{h}^{II}(z) = -z -E_0 - v + \int_{-\infty}^\infty
dE\ {X(E) \over {E+z}} -2\pi i \overline{X}(z), \eqno(4.19)$$
where $\overline{X}(z)$ is a function analytic in the lower half
plane, in a sufficiently large domain, with boundary value
$X(-\lambda)$ on the real axis, for $z \rightarrow \lambda$.
Since, from $(4.19)$, we have the relation
$$ {\rm Im} \overline{h}^{II}(z) = -{\rm Im}z
\biggl( 1 + \int_{-\infty}^\infty dE\ {X(E) \over |E+z|^2}
\biggr) -2 \pi {\rm Re}\overline{X}(z), \eqno(4.20)$$
$\overline{h}^{II}(z)$ can vanish for ${\rm Im}z < 0$.
\par The real part of $\overline{h}^{II}(z)$ is
$$ \eqalign{ {\rm Re}\overline{h}^{II}(z)
 &= -{\rm Re}z \biggl(
1 - \int_{-\infty}^\infty dE\ {X(E) \over |E+z|^2} \biggr)
-E_0 -v \cr &+ \int_{-\infty}^\infty dE\ E {X(E) \over |E+z|^2}
+ 2 \pi {\rm Im}\overline{X}(z) , \cr}\eqno(4.21)$$
and can therefore vanish at the root $\overline\zeta$ for
$$ {\rm Re} \overline\zeta \cong -\omega_0 \cong -(E_0 +v).
\eqno(4.22)$$
{}From $(4.20)$, using the procedure described above to obtain
$(4.17)$, we find that
$$ {\rm Im} \overline\zeta \cong -\pi X(\omega_0), \eqno(4.23)$$
where we have approximated $\overline{X}(\overline\zeta))$ by
its value on the real axis,
 $\overline{X}(-\omega_0) = X(\omega_0)$.  As for $(4.17)$, this
result follows from $(4.20)$ by using the approximation
$$ -{\rm Im}z \int_{-\infty}^\infty dE \ {X(E) \over |E+z|^2}
\cong \pi X(-{\rm Re} z)      \eqno(4.24)$$
for ${\rm Im}z$ small.  Hence the decay rate associated with the
residue of the second pole is the same as that of the first.
\par We wish now to estimate the contour integral $(3.6)$ when the
contour is deformed into the lower half plane \footnote{$\sharp$}
{For large real part, the denominator function $H(z)$ goes
as $O(z^2)$, and the numerator $G(z)$ as $O(z)$.  At the ends
of the long rectangle that brings the integral to the lower half
plane, the contribution of the vertical contour is therefore
bounded by $x^{-1}\int dy\ e^{yt}$, where $y$ is the (negative)
imaginary part of $z$, $x$ is the very large real part of $z$,
and the integral is taken between zero and the distance into
the lower half plane that the contour is carried.  These
contributions therefore vanish for $|x| \rightarrow \infty$,
even for $t \rightarrow 0$.}.
The integral is given by
$$ {1 \over 2 \pi } \int _{-\infty +i\epsilon}^{\infty + i \epsilon}
dz e^{-izt} {\hat A}(z) \cong  \sum_{\rm residues}e^{-izt}
 { 1 \over \overline{h}^{II}(z)h^{II}(z)}
  G^{II}(z) + {\rm background},
\eqno(4.25)$$
where we neglect, for simplicity, the second term of $H^{II}(z)$ in the
denominator, and the background contribution consists of the contour
integral in the lower half plane below the pole contributions; with
a suitable choice of the weight functions, for example, for $X(z),
Y(z)$ Hardy class in the lower half plane on lines with sufficiently large
negative imaginary part, the background contribution can be made
as small as we wish (as in the Pietenpol model$^8$
 in the complex Hilbert
space), given $\epsilon >0$, for $t\geq \epsilon$.
\par To do this, we must also study the analytic continuation
of
$$ \overline{g}(z) \equiv g(-z^*)^* = \int_{-\infty}^\infty
dE\ {Y^*(E) \over {E+z}} + iu^*.  \eqno(4.26)$$
Using the property $Y(-E) = -Y(E)$, it is convenient to rewrite
$(4.26)$ as
$$\overline{g}(z) = \int_{-\infty}^\infty dE {Y^*(E) \over
{E-z}} +iu^* .\eqno(4.27)$$
The analytic continuation is then given, following the procedure
used to obtain $(4.12)$ and $(4.14)$, by
$$\overline{g}^{II}(z) = \int_{-\infty}^\infty dE \
{Y^*(E) \over {E-z}} +iu^* + 2\pi i\overline{Y}(z), \eqno(4.28)$$
where $\overline{Y}(z)$ is a function analytic in the lower
half-plane with the same domain of analyticity as $X(z)$, $Y(z)$,
and with boundary value $Y^*(\lambda)$ in the limit $z
\rightarrow \lambda$ onto the real axis.
At the zero $\zeta$ of $h^{II}(z)$, close to the positive real
axis, near $E=\omega_0$,
$$ \overline{g}^{II}(\zeta) \cong P\int_{-\infty}^\infty
dE\ {Y^*(E) \over {E-\omega_0}} +iu^* +\pi i Y^*(\omega_0).
\eqno(4.29)$$
At the root $\overline\zeta$, close to the negative real
axis, near $E=-\omega_0$,
$$\eqalign{\overline{g}^{II}(\overline\zeta)
 &\cong P\int_{-\infty}^\infty
dE\ {Y^*(E) \over {E+\omega_0}} +iu^* + \pi i Y^*(-\omega_0) \cr
&= \int_{-\infty}^\infty dE\ {Y^*(E) \over {E-\omega_0}}
+iu^*  - \pi i Y^*(\omega_0). \cr} \eqno(4.30)$$
\par We are now in a position to evaluate the pole
contributions to $(4.25)$.
 The residue at $\zeta$ contributes a term which is
approximately
$$ {e^{-i\zeta t} \over
 \overline{h}^{II}(\zeta)} G^{II}(\zeta)=
e^{-i\zeta t} \biggl( 1 +
 {\overline{g}^{II}(\zeta)
 \over \overline{h}^{II}(\zeta)}j\biggr),
\eqno(4.31)$$
where we have neglected small terms in the residue of $h^{II}(z)$.
\par  The zero of the denominator in the
 neighborhood of $z \sim \overline \zeta$ is due
to the factor $\overline{h}^{II}(z)$ in $H^{II}(z)$.  In this case,
however, the first term of $G^{II}(z)$ vanishes, so that the
contribution of the negative energy pole is
$$-e^{i\zeta^* t}  {\overline{g}^{II}(\overline\zeta)
 \over {h}^{II}(\overline\zeta)}j, \eqno(4.32)$$
the quaternionic reflection of the
 resonance below the positive part of the
real line.
The exponential decay rate of both of these contributions
is given by $(4.17)$, i.e., for
$$ \zeta = \omega_0 -i(\gamma / 2), \eqno(4.33)$$
 the decay rate $\gamma$ is given by
$$        \gamma = 2\pi |V_{\omega_0 \alpha}|^2. \eqno(4.34)$$
The oscillation frequencies of the two terms have, however, opposite sign.
\par From $(4.12)$ and $(4.19)$,
 we see that $h^{II}(\overline{\zeta})$
and $\overline{h}^{II}(\zeta)$ are both well-approximated by
$-2\omega_0$; we write $(4.29)$ and $(4.30)$ as
$$ \overline{g}^{II}(\zeta) \cong b
 +\pi i Y^*(\omega_0) \eqno(4.35)$$
and
$$ \overline{g}^{II}(\overline\zeta) \cong
 b - \pi i Y^*(\omega_0) ,\eqno(4.36)$$
where
$$ b = P\int_{-\infty}^\infty dE \
 {Y^*(E) \over {E-\omega_0}} +iu^*. \eqno(4.37)$$
In terms of these quantities,
$$A(t) \cong e^{-i\zeta t} - {1 \over 2\omega_0} \biggl\{
e^{-i\zeta t} (b + \pi i Y^*(\omega_0)) - e^{i\zeta^* t}
(b - \pi i Y^*(\omega_0)) \biggr\} j, \eqno(4.38)$$
and hence
$$\eqalign{|A(t)|^2 &\cong e^{-\gamma t}\biggl\{ 1
+ {1 \over 2\omega_0^2} \bigl[ |b|^2 + \pi^2|Y(\omega_0)|^2 \cr
&-(|b|^2 - \pi^2 |Y(\omega_0)|^2 )\ \cos 2\omega_0 t
- 2{\rm Re}(Y(\omega_0)\pi b \ \sin 2\omega_0 t
 \bigr] \biggr\};\cr}\eqno(4.39)$$
the term in square brackets is non-negative.  It is clear from
$(4.38)$ that in the limit $t \rightarrow 0$, the approximate
expression for $A(t)$ does not approach unity (the approximations
we have made in arriving at this expression do not affect the
quaternionic amplitude in this order). We see, moreover, that
$|A(t)|^2$ approaches a value slightly larger than unity as
$t \rightarrow 0$, in agreement with the usual result$^6$,
implied by the fact that the exact initial decay curve starts
as $1-O(t^2)$ in the neighborhood of $t=0$ (this follows from
the anti-self-adjoint property of $\tilde H$).
  We have argued,
however, that since the background contribution can be made as
small as we wish, the pole contributions must be as accurate
as we wish, for $t >\epsilon \geq 0$.
The non-cancellation of the quaternionic pole residues is
an indication that the series expansion of $A(t)$
 for very small $t$ is not smoothly connected to the
  non-perturbative exponential functions which can represent
$A(t)$ with, in principle, arbitrary accuracy for
non-vanishing $t$.
We shall discuss this
phenomenon a little further in the next section.
\par  The probability
of survival of the unstable state has  small
oscillatory interference terms of twice the frequency ($E_0 + v$),
with  coefficients proportional to the square of $V_{E\alpha}V_{E\beta}$;
 the oscillation in the
survival probability could conceivably provide an observable residual
quaternionic effect. Note that these oscillations do not correspond
to those of the time dependent perturbation expansion for
small $t$ which lead to convergence to the Golden Rule$^2$.
\par If the complex part of $\langle \psi_0 \vert {\tilde V} \vert
\psi_0\rangle$ is large and negative , i.e., $v < -E_0$, the
corresponding complex Hilbert space problem develops a bound state due
to the interaction (the complex pole moves to the left and onto the
negative real axis).  For the quaternionic problem, however, the pole
condition now becomes
$$ \omega_1 =-{\rm Re} \zeta_- \cong E_0 + v.  \eqno(4.40)$$
The pole is, in this case, under
the negative real axis, at $z = \zeta_-$, where (with the
same argument leading to $(4.17)$)
$$ {\rm Im}\zeta_- \cong - \pi
 |V_{\omega_1 \beta}|^2. \eqno(4.41)$$
This very small width corresponds to a resonance that is
an almost-bound state.  The occurrence of a resonance
in quaternionic quantum theory replacing the bound state, in a
corresponding situation, in the complex quantum theory, has been
discussed by Adler$^{1,2}$ both for the one-dimensional and spherically
symmetric three dimensional potential problems.

\par The residue of this pole at $-\omega_1$ contributes
to the survival amplitude a term given approximately by
$$ e^{-i \zeta_- t}\biggl( 1 +
 {\overline{g}^{II}(\zeta_-) \over
\overline{h}^{II}(\zeta_-) }j\biggr),      \eqno(4.42)$$
where $\overline{h}^{II}(\zeta_-) \cong 2\omega_1$.
As in the previous case, there is a second pole, now under the positive
real axis, for which $ -z^* =
 \zeta_-$, i.e., $z= -\zeta_-^* \equiv \overline{\zeta}_-$. The
contribution of this pole to the survival amplitude is
approximately given by
$$ -e^{i\zeta_-^* t}  {\overline{g}^{II}(\overline{\zeta}_-)
 \over h^{II}(\overline{\zeta}_-) }j,
\eqno(4.43)$$
corresponding to a resonance with positive frequency
and very small amplitude, which is the quaternionic
reflection of the almost-bound state on the negative part of the real axis.
\bigskip
\noindent {\bf 5. Conclusions}
\smallskip
\par We have shown that a soluble model for the
description of a quaternionic quantum mechanical system
undergoing decay according the description of Wigner and
Weisskopf$^{10}$ can be constructed in the same way as the
Lee-Friedrichs model$^5$ of the complex quantum theory.
\par The existence of the effective spectrum$^3$
in $(-\infty,\infty)$ of the anti-self-adjoint Hamiltonian
implies that the
problem of the analysis of the complex structure of the amplitude is
quite similar to that of the Pietenpol model$^8$, an idealization of the
Stark Hamiltonian in complex quantum theory. In this model,
 the spectrum of the ``unperturbed'' Hamiltonian has an
absolutely continuous part in $(-\infty, \infty)$, and the
Wigner-Weisskopf theory achieves an almost irreversible law of
evolution (there is no branch point to contribute to long time deviations
from exponential behavior).
\par Carrying out an approximate analysis (assuming the
 discrete-continuous matrix elements of the
 the perturbing potential are
small, and that those of the quaternionic part are
 even smaller) of the analytic properties
of the (analytically continued) amplitudes in the lower half plane, we
find that for perturbations resulting in an energy shift of the real
part that does not become negative, the unperturbed bound state becomes
a resonance of the usual type (although the amplitude has a quaternionic
phase); there is, however,
 a reflection resonance of small amplitude, with
negative frequency, below the negative part of the real axis.  This
reflected resonance results in an oscillatory interference term which
could provide an observable quaternionic residual
effect. If, on the other hand, the energy shift induced by the
diagonal part of the perturbation moves the pole to a position below the
negative part of the real axis, in place of the bound state that would
be found in the corresponding complex theory, one finds a rather narrow
resonance which becomes a bound state in the limit in which the
quaternionic part of the potential vanishes.  Its quaternionic
reflection on
the positive part of the real axis
 is a positive frequency resonance of small
amplitude which vanishes, along with the width of the negative energy
resonance,
 with the quaternionic part of the potential.
 The occurrence, in quaternionic quantum theory, of a resonance
replacing a corresponding bound state
 in the complex quantum theory, has been
discussed by Adler$^{1,2}$ both for the one-dimensional and spherically
symmetric three dimensional potential problems.
\par As we have seen in the previous section, the quaternionic
part of the
 pole residues
at $\zeta$ and $\overline{\zeta}= -\zeta^*$ do not cancel to leave
an approximate unity residue as $t \rightarrow 0$ as one might
expect. The non-cancelling part comes from the behavior of the
quaternionic part $\overline{g}(z)$ of the amplitude near the
real axis, in particular, from a numerator function
(absorptive part) which emerges from the integral due to the
$\delta$-function that forms as $z$ approaches the real axis.
\par To understand this behavior,
 consider the Pietenpol model of the
complex theory$^8$.
  For a Hamiltonian of the form $H=H_0 + V$, where $H_0$ has
continuous spectrum from
 $-\infty$ to $\infty$, and a bound state $\phi$
embedded at $E_0 > 0$, the singularities of the function
$\langle \phi \vert R(z) \vert \phi \rangle$, where $R(z)$ is
the resolvent $(z-H)^{-1}$,
 are determined by
$$ h(z) \langle \phi \vert R(z) \vert \phi \rangle = 1,\eqno(5.1)$$
where
$$h(z) = z -E_0 - v +\int_{-\infty}^\infty {|\langle E\vert V
\vert \phi \rangle |^2 \over {E-z}}, \eqno(5.2)$$
and $v$ is the expectation value of $V$ in the state $\phi$.
The analytic continuation of
 this equation to the lower half plane,
 as for eq.$(4.12)$, has the form
$$ h^{II}(z) = z - E_0 -v +\int_{-\infty}^\infty
dE \ {X(E) \over
{E-z}}  + 2\pi i X(z) , \eqno (5.3)$$
where $X(E) = |\langle E \vert V \vert \phi \rangle|^2$, the
boundary value of the analytic function
$X(z)$ .  Following the arguments
we have given above, one sees that $h^{II}(z)=0$ for
 $$z=\zeta \cong -\pi i X(\omega_0),\eqno(5.4)$$
  for $\omega_0 \cong E_0 + v$. This
result corresponds to a nonperturbative width for the decaying
system. On the other hand, the time dependent
 perturbation expansion  of the amplitude $A(t)$ results in
$$ e^{i(E_0 +v)t} A(t) = 1 - I(t) + {\dots},  \eqno(5.5)$$
where
$$ I(t) = \int _{-\infty}^\infty
 dE\ {X(E)\over (E-E_0 - v)^2}
\bigl\{ i(E-E_0 -v)t +1 - e^{i(E-E_0 -v)t} \bigr\}. \eqno(5.6)$$
It then follows that
$$ \eqalign{|A(t)|^2 &\cong 1 -
 4\int_{-\infty}^\infty dE \ {X(E) \over
(E-E_0-v)^2} \sin^2 \bigl[{(E-E_0-v) \over 2}t \bigr] \cr
&\cong 1 - 2\pi X(E_0+v) t  \cr}, \eqno(5.7)$$
providing a perturbative width in agreement
 with the nonperturbative width obtained in $(5.4)$; the
condition for this result, corresponding to the Golden Rule,
is that $t$ be greater than $1 / E_0$.  However, as we have
pointed out above, the non-perturbative calculation of the
time dependence of the amplitude, as represented by the
pole residues, can be made as accurate as we wish.  The
perturbation expansion, in low order,
 is therefore not an accurate representation
of the decay law, in this case, for any $t > \epsilon \geq 0$.
\par A similar phemonenon occurs for the spectrum of the Stark
effect.  The bound states of an atomic system, such as hydrogen,
are destroyed by the Stark potential (see ref. 8 for a discussion)
for any electric field, no matter how small.  The perturbative
treatment of the energy levels, however, results in a very
good approximation of the level shift from the initially
bound states to the peaks of the spectral enhancements in the
continuous spectrum$^{16}$.
\par  Some development of a formal
scattering theory and the properties of this model for the
description of scattering will be given in a succeeding
publication.
\bigskip
\noindent {\it Acknowledgements. \/} The author is grateful to
S.L. Adler for many stimulating discussions, for his
 very helpful comments on this paper,
 and for his
hospitality at the Institute for Advanced Study,
where this work was supported in part by the
Monell Foundation.
 He also wishes to thank I. Prigogine
and his collaborators in Brussels and Austin, Texas, for many
discussions of the problems of unstable systems.
\bigskip
\vfill
\eject
\noindent {\bf References}
\frenchspacing
\bigskip
\item{1.} S.L. Adler, Phys. Rev. D {\bf 37}, 3654 (1988).
\item{2.} S.L. Adler, {\it Quaternionic Quantum Mechanics}
Oxford University Press, Oxford (1994).
\item{3.} L.P. Horwitz,  Jour. Math. Phys. {\bf 34},3405 (1993).
\item{4.} B. Misra, Proc. Nat. Aca. Sci. {\bf 75}, 1627
(1978); B. Misra, I. Prigogine and M. Courbage, Proc. Nat.
Aca. Sci. {\bf 76}, 4768 (1979).
\item{5.} K.O. Friedrichs, Comm. Pure Appl. Math. {\bf 1},
361 (1950); T.D. Lee, Phys. Rev. {\bf 95}, 1329 (1956).
\item{6.} L.P. Horwitz and J.-P. Marchand, Rocky Mountain
Jour. Math. {\bf 1}, 225 (1973).
\item{7.} L.P. Horwitz and I.M. Sigal, Helv. Phys. Acta
{\bf 51}, 685 (1980); W. Baumgartel, Math. Nachr. {\bf 75}
, 133 (1978); A. Bohm, {\it Quantum Mechanics: Foundations
and Applications}, Springer, Berlin (1986); P. Exner,{\it
Open Systems and Feynman Integrals}, D. Reidel,
Dordrecht (1984); T. Bailey and W. Schieve, Nuovo Cim.
{\bf 47A}, 231 (1978).
\item{8.} See, for example, I. Antoniou, J. Levitan and
L.P. Horwitz, ``Time dependence and intrinsic irreversibility
of the Pietenpol model,'' Institute for Advanced Study preprint
IASSNS 92/59, to be published in Jour. Phys. A:Math. and Gen.
\item{9.} L.P. Horwitz and J. Levitan, Phys. Lett. {\bf A153}
, 413 (1991).
\item{10.} V.F. Weisskopf and E.P. Wigner, Zeits. f. Phys.
{\bf 63}, 54 (1930); {\bf 65},18 (1930).
\item {11.} C.B. Chiu, E.C.G. Sudarshan and B. Misra,
Phys. Rev. {\bf D16}, 520 (1977).
\item{12.} L.P. Horwitz and L.C. Biedenharn, Ann. Phys.
{\bf 157}, 432 (1984).
\item{13.} D. Finkelstein, J.M. Jauch, S. Schiminovich and
D. Speiser, Jour. Math. Phys. {\bf 3}, 207 (1962);
{\bf 4}, 788 (1963).
\item{14.}   See ref. 8 for a discussion of an extension
of the Lee-Friedrichs model for which the unperturbed
Hamiltonian has a continuous spectrum in $(-\infty,\infty)$.
\item{15.} A. Sudbery, Math. Proc. Camb. Phil. Soc. {\bf 85}, 199
(1979); R. Fueter,Comment. Math. Helv. {\bf 7}, 307 (1935).
\item{16.} F. Rellich, {\it\ St\"orungstheorie
 der Spektralzerlegung III\/}, Math. Ann. {\bf 116},555 (1939);
K.O. Friedrichs and P.A. Rejto,
 Comm. Pure and Appl. Math {\bf 15}, 219 (1962).
\vfill
\eject
\end